\documentclass[aps,onecolumn,10pt,final, showpacs]{article}
\usepackage{amsmath,amssymb,amsfonts}
\usepackage{graphics,graphicx}
\usepackage[unicode,colorlinks,linkcolor=blue]{hyperref}
\usepackage[cp1251]{inputenc}   
\usepackage{color}
\usepackage{verbatim}
\usepackage{latexsym}
\usepackage[russian]{babel}

\begin{document}

\title{Baryon number of the Universe as a result \\ of extra space dynamics}

\author{A.V. Grobov\thanks{alexey.grobov@gmail.com} \and S.~G.~Rubin\thanks{sergeirubin@list.ru} \\National Research Nuclear University ``MEPhI''}

\date{}
\maketitle

\abstract{Origin of baryon asymmetry is studied in the framework of extra dimensional approach. Baryon excess production and the symmetrization of extra-space are performed simultaneously. Baryon number is conserved long after the inflationary stage when the U(1) symmetry is achieved.}

\section{Introduction}

Prediction of antimatter made by Dirac in 1928 and
discovery of positrons in cosmic rays initiated the study of antiparticles. Now it is firmly established that our Universe mainly
consists of particles (baryon matter). Antiparticles (antibaryons)
form only a small fraction. The number density of baryons is
characterized by the ratio:
\begin{equation}\label{excess}
\Delta_B = \frac{\Delta n_{B}}{s}= 0.86 \cdot 10^{-10}
\end{equation}
where $s$ - is the entropy density and $\Delta n_{B}$ - is the baryon excess.
This quantity is constant if the baryon number is conserved and the Universe expands adiabatically.

Origin of the baryon asymmetry is one of the key problem in cosmology. The question is whether the Universe was born asymmetric or was the asymmetry formed during
the evolution of the Universe? Why the baryon charge is conserved
today with high accuracy whereas the baryon excess was created in
the past at some (also unclear) stage of the Universe evolution
\cite{Dolgov-obzor}, \cite{1}?

First attempt to explain the phenomenon of baryon asymmetry was done
by Sakharov in 1967. According to his approach there are three known conditions that had to be fulfilled in
order to originate baryon asymmetry at some stage.
Nowadays there are a great variety of different approaches \cite{Fujii}, \cite{Walker}, \cite{Kuzmin},
\cite{Dolgov06}.
Up to now there is no unique preferable model.

As a result there is still no proper answer to the question concerning the emergence of the baryon asymmetry. On the other hand, the idea of
extra space almost inevitably leads to
charge non conservation. Indeed in the framework of multidimensional gravity
observed low energy symmetries are the consequences of isometries of
extra space \cite{Blagojevic}. As was discussed in
\cite{RuSym0,RuSym} there are no any Killing vectors of a nucleated
manifold in the very beginning. The first stage of an extra space
formation consists of its symmetrization. Hence, there are no
conserved charges  at this stage. Charge conservation appears much
later when the extra space geometry acquires
appropriate Killing vectors. The situation is similar to those concerning the problem of baryon asymmetry.

In the present paper we consider a mechanism of the baryon asymmetry
generation accompanied by symmetrization of extra space. It is
assumed that the baryon symmetry (asymmetry) is the consequence of
symmetry (asymmetry) of extra space. The corresponding symmetry of
the theory is restored  during the process of
evolution at later stages.

Another approaches are  known as well. The baryon production due to
dilaton stabilization is discussed in \cite{Dolgov03}. Brane idea
could be fruitful in this context, see \cite{Dvali} where
non conservation of global charge during a
brane nucleation was considered.

In general models based on large and universal extra dimensions have
a serious problem. Indeed, if the true gravity scale is in the TeV
range, early stages of the Universe formation would be seriously
modified \cite{Dolgov06}. In particular it could influence a small
reheating temperature, the amplitude fluctuation of inflaton, the
electroweak symmetry breaking and the particle production during the
reheating stage. One of the way to cure the problem is to take into
account a time variation of the gravitational constant. The latter
should freeze before the nucleogenesis \cite{Bambi,Iocco,Nesseris}.
There is no such problem if an extra space is small enough so that its influence on the inflationary dynamics could be
neglected.

\section{Setup}
Consider a $D=8$-dimensional Riemannian manifold $V_{8}=M_4 \times
V_2 \times H_2$ with a metric $G$. The FRW metric with the scale
factor $a(t)$ of our 4-dim space $M_4$ is denoted as $g_{\mu\nu}(x)$
where $x_{\mu} (\mu =1,2,3,4)$ are its coordinates.  The subspace
$V_2$ with the topology $T_1\times T_1$  possesses a metric
$G^{(V)}_{ab}$ and is described by coordinates  $y_a,  a =5,6$ in
the interval $0\leq y_a <2\pi $. Hyperbolic subspace $H_2$ with a
radius $r_d$ and coordinates $\theta , \phi$ plays a subsidiary role. The interval is chosen in the form
\begin{equation}\label{interval}
ds^2 = dt^2 -dx^2 - dy^2 - dz^2 - (r_{c}^2+h_{ab}) dy_{a}^2 + 2e_{a}dy_{a}d\theta - r_{d}^2 d\theta^2 - r_{d}^2 sinh(\theta)^2 d\phi^2
\end{equation}

As was shortly discussed in the Introduction nucleation of manifolds
containing a symmetry has zero probability \cite{RuSym}. In this
connection consider the metric of the subspace $V_2$ being slightly
deviated from symmetrical one. More definitely, suppose the
following form of the metric
\begin{equation}\label{GV}
G^{(V)}_{ab}=G^{(V,stat)}_{ab}+h_{ab}(t,y_1 ,y_2)
\end{equation}
Its stationary part $G^{(V,stat)}_{ab}=diag(r_c , r_c )$ is
invariant under the $SO(2)$ transformations. The extra space $V_2$
acquires this symmetry at late times when the fluctuations
$h_{ab}(t,y_1, y_2 )$ decay into lighter particles.
Restoration of the symmetry of our metric gives rise to the baryon charge conservation.

The off-diagonal components  of the metric are also important in the
following consideration. $G_{a7}$ metric components are transformed
under the fundamental representation of the $SO(2)$ (or  $U(1)$)
group introduced above. We suppose that the corresponding charge
$Q=Q_B$ is the baryonic one and hence the field $e_a \equiv
G_{a7}$ could form a baryonic condensate. Its interaction with
quarks and leptons will be shortly discussed later in the same way
as it was done in \cite{Dolgov06}.

As will be shown later the baryon asymmetry takes place if
$h_{ab}(t,y_1, y_2 )\neq h_{ab}(t,y_2, y_1 )$. We choose the
simplest form - first term in the Fourier series
\begin{equation}\label{h}
h_{ab}(t,y_1,y_2 ) = \delta_{ab}h(t)\cos(y_1)
\end{equation}
to perform analytical estimation. The $G_{a7}$ components though
small but important.  All other components of the metric $G$ also
contribute to the effective low energy action. Nevertheless we will omit them since they do not contribute to the
baryon excess.  It
strongly facilitates an analysis leaving the idea untouched.

The model is specified by the nonlinear action
\begin{equation}\label{act0}
S=\frac{m_D ^{D-2}}{2} \int d^{D}X \sqrt{G} \left[ R+cR^2 \right]
\end{equation}
There are two parameters in the model - $m_D$ and $c$ while the
metric tensor contains another two - $r_c$ and $r_d$.  According to
modern experiment these extra space sizes must be smaller than $\sim
10^{-18}$ cm. Another restriction followed from the fact that
quantum fluctuations of a metric become important at $m_D$ scale.
The inequalities
\begin{equation}\label{size}
1/m_D \ll r_c , r_d \lesssim 0.1\, \text{TeV}^{-1}
\end{equation}
 permit us to deal with classical behavior of the metric of small enough extra space.

The Ricci scalar $R$ is a complicated function of the fields $h(t,y_1 )$ and $e_a (t)\equiv G_{a7}.$
Keeping in mind expression \eqref{h} the Ricci scalar can be written explicitly:
\begin{equation}\label{Ricciscalar}
R = R_4 + R_{H_2} + R_h + R_e + R_{he}
\end{equation}
where $R_4$ - is a curvature of the 4-dimensional space-time $g_{\mu \nu}$, $R_{H_2}$ is the curvature of the $H_2$ space,
$$R_h = \frac{3}{2 r_{c}^4}\cos^2(y_1) (\partial_{t}h(t))^2 - \frac{1}{r_c ^6}\sin^2(y_1)h(t)^2$$
$$R_e = \frac{1}{2 r_{c}^2 r_{d}^2} (\partial_{t} e_{a})^2 - \frac{2}{r_d ^4 r_c ^2} e_{a}^2 $$
$$R_{he} = -\frac{\coth(\theta)}{r_d ^4 r_c ^6} \sin(y_1) e_{1} h(t) (e_{1}^2+e_{2}^2).$$
To simplify the analysis we will suppose that $r_{d}^2 \gg 4|c|$.

\section{Effective Lagrangian}

After integrating out the internal
coordinates in expression \eqref{act0} one obtains the effective
Lagrangian
\begin{flalign*}
  & S_{eff}=\int d^4 x \sqrt{-g} \mathcal{L}_{eff}, \\
  & \mathcal{L}_{eff}
         = \frac{1}{2}(\partial_{t}\chi)^2 - \frac{1}{2}m^2_{\chi} \chi^2+ \frac{1}{2}(\partial_{t} \varphi_1)^2 - \frac{1}{2}m^2_{\varphi} \varphi_1^2+ \frac{1}{2}(\partial_{t} \varphi_2)^2 - \frac{1}{2}m^2_{\varphi} \varphi_2^2&\\
          &+ c \cdot \lambda^{*}\chi^2 (\varphi_1^2 + \varphi_2^2)^2 \varphi_1^2
\end{flalign*}
where field normalization
\begin{eqnarray} \label{he}
&&\chi= \frac{r_d}{r_c}\sqrt{24\pi^3 m_D ^6 V_{\theta}}\, h\\
&&\varphi_a = \frac{r_d}{r_c}\sqrt{8\pi^3 m_D ^6 V_{\theta}}\,e_a \\
&& V_{\theta}  = \int_{\phi_{-}}^{\phi_{+}} \int_{\theta_{-}}^{\theta_{+}} \! \sinh(\theta) \mathrm{d} \theta  \mathrm{d} \phi
\end{eqnarray}
was performed to obtain the standard form of the Lagrangian. Here we
omitted terms describing the Einstein-Hilbert
action and $\Lambda - $ term because they do not play significant role in the model.
To obtain them accurately, one should take into account all effects what is far
from our purpose. Besides we deleted all interaction terms except the last one
which is responsible for the baryogenesis. It can be done if the
metric fluctuations are small
\begin{equation}\label{ineq2}
h(t,y_1, y_2 )\ll r_c ^2 ,\quad e_a \ll r_c ^2 , r_d ^2 .
\end{equation}
The masses of the new fields are expressed in terms of initial parameters
\begin{equation}\label{masses}
m^2_{\chi} = \frac{1}{3r_c ^2},\quad m_{\varphi} ^2 = \frac{4}{r_d ^2}
\end{equation}
The expression for coupling constant
\begin{equation}\label{lambda}
\lambda^{*} = \frac{W_{\theta} / V_{\theta}^4}{3072\pi^9 m_D ^{18} r_c ^8 (r_d ^2 - 4c)^4}
\end{equation}
contains the integral
\begin{equation}
W_{\theta} = \int_{\phi_{-}}^{\phi_{+}} \int_{\theta_{-}}^{\theta_{+}} \cosh(\theta)\coth(\theta) \mathrm{d} \theta \mathrm{d} \phi
\end{equation}
over the space $H_2$ that are worth discussing.


The limits of integration are nontrivial due to complicated boundary of compact
hyperbolic manifolds. The boundary of our compact space $H_2$ with
the metric
\begin{equation}
ds^2 = d \theta^2 + sinh^{2}(\theta) d \phi^2
\end{equation}
is described in the following way
\begin{equation} \nonumber
\phi_{-} = 0, \quad \phi_{+} = \frac{\pi}{4}
\end{equation}
\begin{equation} \nonumber
\ \theta_{+} = 2arth \left( A \cdot ctg(\frac{\pi}{8}) cos(\phi - \frac{\pi}{8}) - \sqrt{A^2 cos^{2}(\phi - \frac{\pi}{8}) - sin^{2}(\phi - \frac{\pi}{8})} \right)
\end{equation}
where $A = 2^{\frac{1}{4}}sin(\frac{\pi}{8})$. The lower limit
$\theta_-$ is worth  discussing. From a pure geometrical point of
view $\theta_- =0$. On the other hand a classical description is valid if size of the space is much greater than the Planck scale. Smaller regions can not be considered accurately due to strong quantum fluctuations inside them.
So the scale of lower limit $\theta_-$ of action integral
is proportional to $1/M_{Planck}$. Usually, such a small value does
not influence any effects. In our case the quantum fluctuations of
the metric became important at the scale $1/m_D$ instead of
$1/M_{Planck}$. So the value of the low limit of the integral over the angle
$\theta$ should be $\theta_- \simeq 1/(r_d m_D)$. After integration
$V_{\theta} \approx 1.57$, and $W_{\theta}\simeq ln(m_D r_d)$.

In terms of complex field $$\phi = \frac{\varphi_1 + i
\varphi_2}{\sqrt{2}}$$ the effective Lagrangian has the form (we
assume $m^2 _{\varphi} \equiv m^2 _{\phi}$):
\begin{eqnarray}\label{lagreff}
  \mathcal{L}_{eff} &=& \frac{1}{2}(\partial_{t}\chi)^2 - \frac{1}{2}m^2_{\chi} \chi^2+ \partial_{t} \phi \partial_{t} \phi^{*} - m^2 _{\phi} \phi \phi^{*}\\
  &-& 2\lambda\chi^2 \phi^{2}\phi^{*2} \left[\phi+\phi^{*}\right]^2 \nonumber
\end{eqnarray}
where  $\lambda = -c \cdot \lambda^{*} >0$ and $c<0$. The last term
breaks global $U(1)$ symmetry and therefore is responsible for
asymmetrical baryosynthesis. In the modern epoch $\chi (t\rightarrow
\infty)\rightarrow 0$ so that the $U(1)$ is restored. The latter
group is isomorphic to the $SO(2)$ group - exact symmetry group of
unperturbed metric $G^V$ in \eqref{GV}. Thus the process of
symmetrization of the extra space $V_2$ is accompanied by the baryon
excess production.

The final form of the effective action
\begin{equation}\label{actfin}
S_{r, \vartheta} = \int dt a^3(t)\left[ \frac{1}{2}(\partial_{t}\chi)^2 - \frac{1}{2}m^2_{\chi} \chi^2 + \frac{1}{2} \dot{r}^2 + \frac{1}{2}r^2 \dot{\vartheta}^2 - V(r, \vartheta, \chi)\right]
\end{equation}
contains the potential
$$V(r,\vartheta, \chi) =  \frac{m_{\phi} ^2 r^2}{2} + \lambda \chi^2 r^6 \cos^2(\vartheta)$$
Here the field $\phi$ is represented in the form
$
\phi(t) = r(t)e^{i\vartheta(t)}/\sqrt{2}
$

Effective Lagrangian \eqref{lagreff} is similar to those considered
in the framework of Affleck-Dine model  \cite{AD}. The only
essential difference is the presence of additional field $\chi$. The
baryosynthesis is terminated when this field reaches zero value.

\section{Baryon excess}
In the FRW space the equations of motion for the fields $\chi , r$ and $\vartheta$
\begin{eqnarray}
&&\ddot{\chi}+3H\dot{\chi} +m^2 _{\chi} \chi = -2\lambda \chi r^6 cos^2(\vartheta) \nonumber \\
&&\ddot{r}+3H\dot{r}-r\dot{\vartheta}^2 + m^2 _{\phi} r = -6\lambda \chi^2 r^5 \cos^2(\vartheta) \nonumber \\
&& r^2 \ddot{\vartheta}+3Hr^2 \dot{\vartheta} + 2r\dot{\vartheta}\dot{r} = \lambda \chi^2 r^6 \sin (2 \vartheta) \label{17}
\end{eqnarray}
follow from action \eqref{actfin}.

Oscillations of the field $\vartheta$ are responsible for
the generation of the baryon excess. Indeed the dynamical equation for the field $\vartheta$ can be written in more suitable form
\begin{equation}\label{eqmain}
\frac{1}{a^3} \frac{\partial}{\partial t} \left( a^3 r^2 \dot{\vartheta}\right) = -\frac{\partial V}{\partial \vartheta}
\end{equation}
The field $\phi$ is transformed  under the
fundamental representation of the group $U(1)$. The baryon charge
connected to this group is calculated in standard manner $n_B = j_0
=i(\phi^*\partial_{0}\phi - \phi \partial_{0}\phi^*)= r^2
\dot{\vartheta}$. Substituting this into equation \eqref{eqmain} we
obtain the equation for baryon density
\begin{eqnarray}
&&a^{-3}\frac{\partial}{\partial t}(a^3 n_B)=
-\frac{\partial V}{\partial \vartheta}=\lambda \chi^2 r^6 sin(2\vartheta)
\label{thetaclass}
\end{eqnarray}
with formal solution
\begin{equation}\label{nB}
n_B (t)=  a(t)^{-3}\lambda \int_{t_{in}}^t a(t')^3 r(t')^6  \chi(t')^2 sin(2\vartheta(t'))dt' .
\end{equation}

Our aim is to study the ability of the model to explain
the observable baryon density. According to \cite{Rubakov} it can be
done in quite simple and elegant way. Suppose that the dynamic of the field
$\vartheta$ ruled by equation \eqref{eqmain} elaborates baryon excess during one e-fold. Then the estimation of integral \eqref{nB} gives
\begin{equation}\label{nBa}
n_B (t_B )\simeq  e^{-3}\lambda \chi^2 sin(2\vartheta) r^6 H ^{-1}.
\end{equation}
A baryon charge at the moment  $t_B = H ^{-1} $ of its creation is
connected to the modern baryon excess $n_B(t_0)$
\begin{equation}\label{nBb}
n_B(t_B ) =(a(t_B )/a(t_0 ))^3 n_B(t_0)=(t_0 H )^{2}n_B(t_0).
\end{equation}
Here we suppose the approximate equality  $a(t_B)/a(t_0 )\simeq (t_0
/t_B )^{2/3}\simeq(t_0 H )^{2/3}$ to simplify the estimation.

In this paper we suppose the masses of the field $\chi$ and $r$ \eqref{masses} is of order $10^{12}$ GeV for chosen set of the parameters, see the text above expression  \eqref{Hrange},  what is much greater than the Hubble parameter in the considered stage. It means that these fields quickly oscillate around zero. The periods of these oscillations are much smaller than the period of oscillations of the
field $\vartheta$ what allows us to use average values like $\chi^2
\rightarrow \langle\chi^2\rangle$ in expression \eqref{nBa}. Keeping
in mind expression \eqref{nBb} we obtain the connection between the
parameters
\begin{equation}
n_B(t_0) = t_0^{-2}H ^{-3} e^{-3}\lambda \langle\chi^2\rangle \langle r^6\rangle sin(2\vartheta)
\end{equation}
The observed parameters included in this equation are $
 t_0=14 \cdot 10^{16} sec=6.3 \cdot 10^{41}\, \text{GeV}^{-1},
$
$
 n_B = 2.46 \cdot 10^{-7}cm^{-3} = 1.9 \cdot 10^{-50}\, \text{GeV}^3 .
$
Finally, the relation between unknown parameters acquires the form (we suppose $sin(2\vartheta) \approx 1$)
 \begin{equation}\label{eq1}
\frac{\lambda \langle\chi^2\rangle \langle r^6\rangle}{H ^3} \approx 3.8 \cdot 10^{36}\, \text{GeV}
 \end{equation}

Effective baryon production starts when a slow rolling of the field
$\vartheta$ is terminated what leads to an additional
connection
\begin{equation}\label{eq2}
3H\simeq \sqrt{\lambda \langle\chi^2\rangle \langle r^4\rangle}
\end{equation}
followed from the third equation in \eqref{17}. The estimations could be violated at high energies by forth order terms which have been omitted in action \eqref{actfin}. We assume that the process of baryon formation is most effective at moderate energies where the quadratic terms dominate.  At high energies and large field values quick expansion of the Universe strongly reduces the baryon number density.
With estimation of the integral $W_{\theta} \approx ln(m_D r_d)$ $\lambda$ can be written as
 \begin{equation} \label{lmbd1}
 \lambda \approx -1.8 \cdot 10^{-9}\frac{c}{ m_D ^{18} r_c ^8 r_d ^8} ln(m_D r_d)
 \end{equation}
Expressions \eqref{eq1} and \eqref{eq2} impose main restrictions on the parameters of the model.

Let us specify parameters of the model and choose $m_D =10^{14}$
GeV,$c= - 42 m_{D}^{-2}$, $r_c = r_d =10^2 m_{D}^{-1}$.
Initial values of the fields $h$ and $e_a$ should be small compared to  the size of
extra space \eqref{ineq2} and we assume \eqref{he} $\langle |\chi| \rangle \sim
\langle |\phi| \rangle \sim 10^{6} m_{D}$. This set of parameters
satisfies main equations \eqref{eq1} and \eqref{eq2} if the Hubble
parameter $H\sim 2.4 \cdot 10^4$ GeV.
The Hubble parameter lays in a wide range
 \begin{equation}\label{Hrange}
 0.1\, \text{GeV} < H < 10^{13}\, \text{GeV}
 \end{equation}
if the process takes place after the inflation and before the
primordial nucleosynthesis and our choice seems reasonable.

In addition, there are several natural relations between the parameters like those represented in \eqref{size} and \eqref{ineq2}. It can be easily checked that all of them are satisfied for chosen values of the parameters. The energy density of the fields $\phi$ and $\chi$ is much smaller than the energy density of the inflaton field. Indeed
 \begin{equation} \label{chifieldenergy}
\rho_{\chi} \simeq \frac{1}{2}m_{\chi}^2 \chi^2 \sim  10^6 \cdot  m_{D}^4  = 10^{-14} M_{pl}^4.
 \end{equation}
At the same time the inflaton energy density is about $\rho_{inf} = 10^{-12} M_{pl}^4$ in the framework of the chaotic inflation.

The relations mentioned above constrain the permissible range of parameters that is represented in Fig. 1. One can see that there is substantial room for the parameters inside the triangle. Typically, they could vary in order of magnitude so that there is no fine tuning in this model.
\begin{figure}[h!]\label{Parameterspace}
\begin{center}
\includegraphics[scale=0.4]{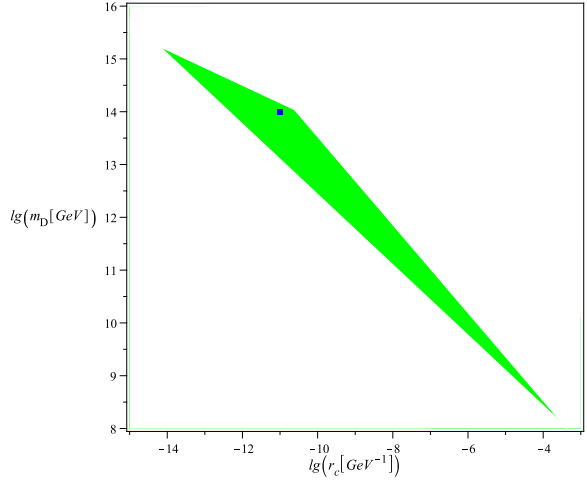}
\caption{The gray triangle represents the parameter space in terms of two parameters $m_D$ and $r_c$. Parameters used in the text are marked as the small square.}
\end{center}
\end{figure}

In our approach we suppose that the observed symmetries are the result of the appropriate symmetries of an extra space. It is known that the internal $SU(2)\times U(1)$ electroweak symmetry of the Standard Model could exist due to some symmetry of an extra space \cite{BOBRU}, \cite{Montani}. The conservation of the baryon and lepton charges is supposed to be the result of some $U(1)$ symmetries of extra space as well. In this paper we discuss the $U(1)_B$ symmetry of the extra space $V_2$ which is responsible for the baryon charge conservation. Those fermions transformed under fundamental representation of the  $U(1)_B$ group are observed as baryons.
The baryon charge stored by the field $\phi$ should be transferred to matter fields. As was shown in \cite{Dolgov-obzor,Dolgov94} it can be done due to the interaction term of the form
$$L_{int} = g\phi \bar{Q} L + h.c.,$$
where $Q$ is a wave function of some hypothetical heavy quark, $L$ is a wave function of lepton and $g$ is a Yukawa coupling constant. 

The coupling constant $g$ should be quite small for not to destroy the model developed above. Otherwise, quanta of the field $\phi$  would have decayed into fermions before its condensate starts to oscillate after the moment $t_{osc}\sim H^{-1}$. The condensate of the field $\phi$ is evaporated during the time $t_{decay}\sim \Gamma_{\phi}^{-1}$ where $\Gamma_{\phi}$ is the decay probability. As was discussed in \cite{DolgKir} its form is
\begin{equation} \label{decay}
\Gamma_{\phi} = 4 \pi^{-5/2}\frac{g^2 m_{\phi}}{4\pi} \left( \frac{m_{\phi}}{g\langle |\phi| \rangle} \right)^{1/2} 
\end{equation}
if $g\langle |\phi| \rangle > m_{\phi}$. This inequality is true for the parameter values chosen above. As the result the inequality
\begin{equation}
t_{osc}< t_{decay}
\end{equation}
gives the upper limit to the fermion coupling constant $g\leq 2.5 \cdot 10^{-2}$. Accordingly, upper limit for effective coupling constant $g_{eff}\equiv g^2 /(4\pi)$ is approximately $0.5\cdot 10^{-4}$.  This limit looks reasonable.

If the symmetry groups discussed above are local ones the model should contain gauge fields.  Gauge fields associated with baryon/lepton charge are discussed in a literature, see e.g. \cite{Lebed}. In our case gauge fields should be represented by off diagonal components of metric tensor (\ref{interval}). Nonobservation of baryon gauge field could be explained by a large mass of its quanta. This subject worths separate paper.

\section{Conclusion}

Among others there are two key problems  - the observable baryon
excess in the Universe and the existence of various
symmetries. The latter is tightly connected to appropriate symmetries of an extra space in the framework of extra dimensional
approach. So the second problem may be reformulated as the question
"Why an extra space possesses any symmetries?"\,.  If it was born having an arbitrary geometry, there must exist
some period during which the symmetries are
established. This means that no charges are conserved during this
stage. In the following stage the extra space is stabilized and acquires
some symmetries. This
leads to charges conservation in the modern epoch.

The picture described above fits very well with the baryon problem. Non conservation of the baryon charge results in the baryon excess in the beginning. In the
modern epoch an extra space is supposed to be stable and possesses $U(1)$ symmetry which relates to conservation of baryon charge. In this paper we show that this mechanism is able to explain the observable baryon excess.

\section{Acknowledgment}
 S.R. was supported by the Ministry of education and science of Russian Federation, project 14.A18.21.0789. The work of A.G. was supported by The Ministry of education and science of Russian Federation, project 14.132.21.1446. The authors are grateful to A. A. Korotkevich for clarifying the problem concerning a boundary of compact hyperbolic space.

\end{document}